%% file: main.tex
\documentclass[twocolumn,prd, aps, notitlepage, nofootinbib, showpacs, superscriptaddress,longbibliography]{revtex4-1}
\usepackage[utf8]{inputenc}
\usepackage{graphicx,mathtools,xcolor,latexsym,rotating,soul}
\usepackage[T1]{fontenc}
\usepackage{tikz}
\usepackage{comment}
\usepackage{amsmath, amssymb, amsthm, amsfonts,breqn}
\usepackage{multirow}
\usetikzlibrary{decorations.pathmorphing}
\usepackage[colorlinks=true,citecolor=green,urlcolor=blue,linkcolor=blue]{hyperref}
\usepackage{pifont}
\usepackage{gensymb}
\usepackage{enumitem}
\usepackage{mathrsfs}

\newtheoremstyle{named}{}{}{\itshape}{}{\bfseries}{.}{.5em}{\thmnote{#3 }#1}
\theoremstyle{named}

\newcommand*{\rom}[1]{\expandafter\@slowromancap\romannumeral #1@}

\newcommand{\stf}[1]{\langle #1 \rangle}

\makeatletter
\let\cat@comma@active\@empty
\makeatother
\begin{document}
\title{Dissipative tidal effects to next-to-leading order and constraints on the dissipative tidal deformability using gravitational wave data}

\author{Abhishek Hegade K. R.}
\email{ah30@illinois.edu}
\affiliation{Illinois Center for Advanced Studies of the Universe, Department of Physics, University of Illinois Urbana-Champaign, Urbana, IL 61801, USA}

\author{Justin L. Ripley}
\email{ripley@illinois.edu}
\affiliation{Illinois Center for Advanced Studies of the Universe, Department of Physics, University of Illinois Urbana-Champaign, Urbana, IL 61801, USA}

\author{Nicol\'as Yunes}
\email{nyunes@illinois.edu}
\affiliation{Illinois Center for Advanced Studies of the Universe, Department of Physics, University of Illinois Urbana-Champaign, Urbana, IL 61801, USA}
\begin{abstract}
Dissipative tidal interactions can be used to probe the out-of-equilibrium physics of neutron stars using gravitational wave observations.
In this paper, we present the first post-Newtonian (PN) corrections to the orbital dynamics of a binary system containing objects whose tidal interactions have a dissipative contribution.
We derive the 1PN-accurate equations of motion in the center-of-mass frame and a generalized energy-balance law that is valid for dissipative tidal interactions.
We show how mass and energy loss due to the absorption of orbital energy change the orbital dynamics and derive the next-to-leading order correction to the gravitational wave phase of a binary system in a quasi-circular orbit containing initially non-spinning components.
We then use this waveform model to constrain, for the first time, the individual dissipative tidal deformabilities of each of the binary components that generated the GW170817 event using real data. 
We find that the GW170817 data requires $\Xi_{1} \lesssim 1121$ and $\Xi_{2} \lesssim 1692$ at 90\% confidence, where $\Xi_{1,2}$ are the individual tidal deformabilities of the primary and secondary binary components that produced the GW170817 event. 
\end{abstract}
\maketitle
\allowdisplaybreaks[4]
\section{Introduction}
A neutron star in a binary system is tidally deformed by its companion, and the strength of the tidal deformation depends on the internal properties of the star~\cite{Flanagan:2007ix}.
The tidal deformation affects the gravitational wave phase and helps one probe the properties of high density nuclear matter, an important unsolved problem in astrophysics and nuclear physics~\cite{Ozel:2016oaf,Lattimer:2021emm,Burgio:2021vgk}.
The tidal deformation of a neutron star depends both on its equilibrium properties, such as the equation of state~\cite{Hinderer_2010}, and on its out-of-equilibrium properties, such as the internal viscosity of the nuclear matter in neutron stars or the absorption of gravitational waves by the event horizon of a black hole~\cite{Ripley_2023,Tagoshi_1997}.

Tidal effects that conserve the orbital energy of the binary correct the gravitational wave phase of point particles starting at 5PN order~\cite{Flanagan:2007ix}. 
A great amount of work has been carried out to model the conservative tidal response of compact objects~\cite{Hinderer_2008,Damour:2009vw,Binnington:2009bb}.
Data from the gravitational wave event GW170817~\cite{LIGOScientific:2018hze} has also been used to constrain the 5PN term in the gravitational wave phase and understand the properties of the equation of state of high density nuclear matter~\cite{LIGOScientific:2018cki}. 
The constraints on the conservative tidal effects are expected to improve with next-generation detectors, leading to more stringent inferences on the equation of state~\cite{Chatziioannou_2020}.

For conservative tidal interactions, the leading (5PN order) tidal contribution to the gravitational wave phase depends on an effective combination of the tidal deformabilities of each star~\cite{Flanagan:2007ix}.
To break the degeneracy in this effective combination and measure the individual tidal deformabilities of each star in the system, it is important to find PN corrections to the leading PN order contribution.
Motivated by these considerations, conservative tidal effects have been calculated to 6PN order in the gravitational wave phase~\cite{Vines_Waveform_2011,Bini-Damour-Faye} for quadrupolar tidal effects. 
The current state-of-the-art model incorporates electric quadrupolar, electric octupolar and current quadrupolar tidal effects to next-to-next-to leading order in the gravitational wave phase~\cite{Henry_2020_eom,Henry_2020_phase}. 
Several studies have also considered the effects of rotation in the conservative tidal interaction~\cite{Abdelsalhin_2018,landry2018rotationaltidal,Gupta_2021} and dynamic contributions due to resonant excitation of fluid modes inside the neutron stars~\cite{Hinderer:2016eia,Pratten:2021pro}.

Tidal effects that dissipate the orbital energy of the binary correct the gravitational wave phase of point particles, starting at 4PN order for non-spinning objects~\cite{Poisson_1995,Ripley_2023} and at 2.5PN order for spinning objects~\cite{Tagoshi_1997}.
For black holes, the source of dissipation is the absorption of gravitational waves by the event horizon~\cite{Poisson_1995,Tagoshi_1997}.
For neutron stars, the source of dissipation is internal mechanisms, such as bulk or shear viscosity of the high density nuclear matter~\cite{Ripley_2023,HegadeKR:2024agt}.
To calculate the dissipative tidal response of black holes and exotic compact objects, techniques from black hole perturbation theory (BHPT)~\cite{Poisson_2004,Chia:2020yla,Saketh:2023bul,Chakraborty:2023zed}, PN theory~\cite{Poisson_2004,Taylor_2008,Poisson:2020vap} and effective field theory~\cite{Chia:2024bwc} have been used.
The first calculation of the dissipative tidal response of non-spinning neutron stars was carried out in~\cite{HegadeKR:2024agt}, using a polytropic equation of state for the nuclear matter.
The first constraints on the leading PN order dissipative tidal deformability were then obtained in~\cite{Ripley:2023lsq} using the GW170817 event. Since these constraints probe the internal dissipative mechanisms of neutron stars, they open a new avenue to probe the out-of-equilibrium properties of high density nuclear matter with gravitational waves.

Just as in the case of the conservative tidal deformabilities, in order to break the degeneracy between the individual dissipative tidal deformabilities of each star it is necessary to go beyond leading PN order in the calculation of the gravitational wave phase.
For black holes, these PN corrections have been calculated using several methods.
The first calculations were carried out in~\cite{Tagoshi_1997} using black hole perturbation theory in the extreme mass-ratio limit, which focused on the black hole energy absorption rate and was completed to very high PN orders~\cite{Tagoshi_1997}.
For the case of comparable-mass spinning black holes in binary systems, corrections to the gravitational wave phase up to 1.5PN order higher than the leading 2.5PN effect were calculated in~\cite{Taylor_2008,Chatziioannou_2016,Saketh:2023bul}.
While important, these spin-dependent effects are suppressed for neutron stars, because their spin is expected to be small. 
Therefore, to break the degeneracy between the dissipative tidal deformabilities of each neutron star in a binary, it is necessary to calculate the next-to-leading PN order correction to the 4PN effect, which is the main focus of this paper.

We accomplish this goal in this paper by leveraging important previous results in the PN literature.  
Racine and Flanagan derived the 1PN correction to the equations of motion for compact objects with arbitrary internal structure in~\cite{Racine:2004xg,Racine_2005}, generalizing the work of Damour, Soffel and Xu (DSX)~\cite{DSX-I,DSX-II,DSX-III}. 
Later, Vines and Flanagan (VF) used the formalism of Racine and Flanagan to derive the 1PN correction to the equation of motion of binary systems, incorporating quadrupolar and spin-orbit interactions~\cite{Vines_flanagan_2010}.
In this paper, we specialize the formalism of VF to the case of dissipative quadrupolar interactions and derive the effect of tidal dissipation on the orbital motion of the binary system in the center of mass (CoM) frame to 1PN order.
In the presence of dissipative tidal interactions, the mass and spin of the objects in the binary are not conserved.
We derive a generalized energy balance equation and showcase how dissipative tidal interactions remove energy from the orbit and how the evolution of the mass and spin of the object contribute to the dissipative tidal flux.
We then specialize our discussion to the case of initially non-spinning compact objects in a quasi-circular binary inspiral and present the 5PN correction (relative to the leading PN order point-particle contribution) due to dissipative tidal interactions in the gravitational wave phase.

The first constraint on the dissipative tidal deformability of neutron stars was obtained by analyzing the GW170817 event with the leading PN order dissipative tidal waveform~\cite{Ripley:2023lsq}, which depends on a certain combination of the individual dissipative tidal deformabilities. 
Because of this, the individual dissipative tidal deformabilities of the binary components that produced the GW170817 event could not be constrained using that model due to correlations between the individual deformabilities. 
Due to these degeneracies, only heuristic estimates on the magnitude of the individual dissipative tidal deformabilities were made in~\cite{Ripley:2023lsq}, by assuming that the binary components that produced the GW170817 event had exactly the same mass and the equation of state.
Here, we improve on these heuristic calculations by using the GW170817 data to properly estimate and constrain the individual dissipative tidal deformabilities through a Bayesian analysis that uses the next-to-leading PN order dissipative tidal waveform model that we have calculated. 
While we find that the GW170817 data is not sufficiently informative to measure the individual dissipative tidal deformabilities, we do place the first constraints on these quantities for neutron stars. More precisely, we find that $\Xi_{1} \lesssim 1121$ and $\Xi_{2} \lesssim 1692$ to 90\% confidence, where $\Xi_{1,2}$ are the individual tidal deformabilities of the primary and secondary binary components that produced the GW170817 event. These constraints will enable us to bound the magnitude of the bulk or shear viscosities inside each star in this binary, once more detailed theoretical calculations are carried out to connect these viscosities to the dissipative tidal deformabilities. 

The rest of this paper presents the details of the calculations summarized above and it is outlined as follows. In Sec.~\ref{sec:EoM} we provide a summary of the 1PN equations of motion presented by VF and we derive the generalized energy-balance equation for dissipative tidal interactions.
In Sec.~\ref{sec:GW-phase}, we then derive the gravitational wave phase for a binary system in a quasi-circular orbit containing non-spinning objects.
In Sec.~\ref{sec:GW170817}, we analyze the GW170817 event using the gravitational phase derived in Sec.~\ref{sec:GW-phase}. 
In Sec.~\ref{sec:conclusions}, we present our conclusions and discuss possible future research. 
Henceforth, we use the following conventions: 
the signature of our metric is $(-,+,+,+)$; 
we set $G=1$ but retain powers of $c$ in our expressions to count PN orders; 
spacetime indices are labeled with lower-case Greek letters $(\alpha,\beta,\ldots)$, while spatial indices are labeled lower-case Latin letters $(a,b,c,\ldots)$ in the middle of the alphabet;
we use $\stf{\cdots}$ in index lists to denote the symmetric trace-free combination of tensorial indices;
repeated indices stand for the Einstein summation convention, where we raise and lower indices with the flat spacetime metric (because we are working in PN theory).  

\section{Equations of motion}\label{sec:EoM}

In this section, we summarize the 1PN equations of motion, following primarily the work of VF in~\cite{Vines_flanagan_2010}, and we calculate the generalized energy-balance equation for binaries with dissipative tidal interactions to 1PN order. In particular, we setup notation in Sec.~\ref{sec:notation}, and calculate the 1PN equations of motion and the energy flux due to dissipative tidal interactions in Sec.~\ref{subsec:EoM}.
We describe the tidal response function in Sec.~\ref{sec:tidal-response} and derive the generalized energy balance equation in Sec.~\ref{sec:generalized-lagrangian}.
\begin{figure}[t!]
    \centering
    \includegraphics[width = 1 \columnwidth ]{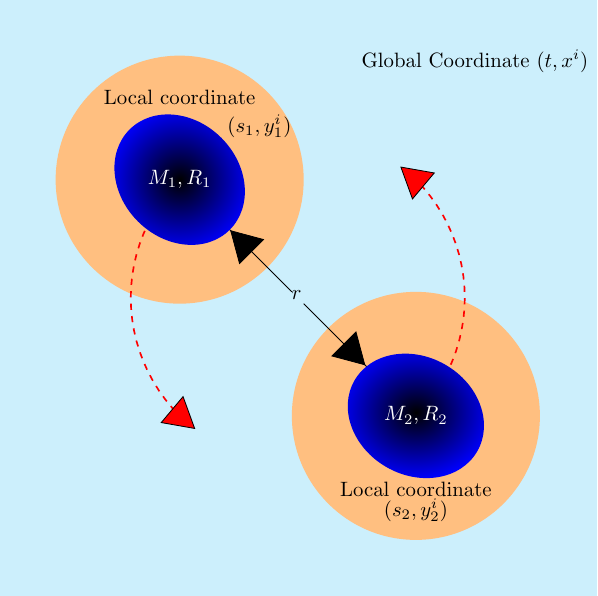}
    \caption{Cartoon (not to scale) depicting the motion of two tidally interacting neutron stars (shown in dark blue, with masses $M_{1,2}$ and radii $R_{1,2}$) in a quasi-circular orbit of radius $r$ on a constant time slice.
    The spacetime is separated into three zones: the inner body zones (shown in orange, close to objects $1$ and $2$), where we employ a local coordinate systems $(s_{1,2},y^i_{1,2})$, and a post-Newtonian zone (shown in light blue, far from either object), where we employ a global coordinate system $(t,x^i)$.
    }
    \label{fig:zones-cartoon}
\end{figure}
\subsection{Notation}\label{sec:notation}
We follow the notation of~\cite{Vines_flanagan_2010,Vines_Waveform_2011} to describe the orbital equations of motion and the gravitational wave phase.
We warn the reader that this notation is different from that used in our earlier paper~\cite{Ripley_2023}, which derived the leading PN order phase contribution at 4PN order.
We choose the notation of~\cite{Vines_flanagan_2010,Vines_Waveform_2011} to make comparisons easier.

We label the two objects in the binary, and any physical
quantities associated with them, with the subscripts $1$ and $2$. 
We use the word ``object'' instead of ``neutron star'' or ``black hole'', to remain agnostic about the system in question.
The masses of the two objects are $M_1$ and $M_2$, the total mass is $M\equiv M_1+M_2$, the reduced mass is $\mu \equiv M_1 M_2/M$, and the symmetric mass ratio is $\eta\equiv \mu/M$.
The characteristic radii of the objects are $R_1$ and $R_2$, which could refer to the equatorial radius for a neutron star or the areal radius for a black hole, while their orbital separation is denoted by $r$.
We also define the mass ratio $q = M_2/M_1$ and\footnote{Note that VF use $\chi_{1,2}$ instead of $X_{1,2}$.} 
$X_{1,2} \equiv M_{1,2}/M$ with the convention that $M_{1} > M_{2}$.
The compactness of each body is defined by $C_{1,2} \equiv {M_{1,2}}/({R_{1,2} c^2})$.
Following VF, we define three coordinate systems: a global conformally harmonic coordinate system denoted by $\left(t, x^j \right)$, and two local coordinate systems covering the neighborhood of the two objects in the binary denoted by $\left(s_{1,2}, y_{1,2}^j \right)$. 
A schematic diagram of such a binary system is presented in Fig.~\ref{fig:zones-cartoon}. 

We denote the (electric-type, quadrupolar) dissipative tidal deformability of each object by $\Xi_{1,2}$.
The dissipative tidal deformability quantifies the dissipative tidal response of the object to the external tidal field. Schematically, the dissipative quadrupolar deformation of object $A$ when in presence of an external time-dependent quadrupolar tidal field is
\begin{equation}
Q^{ab, {\rm diss}}_{A} = - \frac{\Xi_{A}}{c^{13}} M_{A}^{6} \frac{d}{d s_{A}} G^{ab} \,,
\end{equation}
where $Q_{ab}^{{\rm diss}}$ is the dissipative contribution to the (mass-type) quadrupole field tensor and $G_{ab}$ is the (electric-type) quadrupole tidal field tensor (see Eq.~\eqref{eq:Q2-diss-eqn} below for more details).
We also define two combinations of the individual dissipative tidal deformabilities, which we refer to as the binary dissipative tidal deformabilities, namely 
\begin{subequations}\label{eq:xibar-delta-xibar}
\begin{align}
    \label{eq:bar-Xi}
    \bar{\Xi} &
    \equiv
    f_1(\eta) \frac{\Xi_1 + \Xi_2}{2}
    +
    g_1(\eta) \frac{\Xi_2 - \Xi_1}{2} 
    \,,\\
    \label{eq:bar-delta-Xi}
    \delta \bar{\Xi} & \equiv
    f_2(\eta) \frac{\Xi_1 + \Xi_2}{2}
    +
    g_2(\eta) \frac{\Xi_2 - \Xi_1}{2}
    \,,
\end{align}
\end{subequations}
where
\begin{subequations}
\begin{align}
    f_1(\eta)
    &= 
    8\left(2 \eta^2-4 \eta +1\right) 
    \,,
    \\
    g_1(\eta) 
    &=
    -8\sqrt{1-4 \eta } (1-2\eta)
    \,,
    \\
    f_2(\eta)
    &=  
    \frac{1568 \eta^3}{753}+\frac{3288 \eta^2}{251}-\frac{23536 \eta}{753}+\frac{5996}{753}
    \,,
    \\
    g_2(\eta) 
    &=
    \sqrt{1-4\eta} \bigg[\frac{1232 \eta^2}{753}+\frac{3848 \eta}{251}-\frac{5996}{753} \bigg]
    \,.
\end{align}
\end{subequations}
The binary dissipative tidal deformabilities
will appear in the gravitational waveform that we calculate in this paper.

Following VF, we define an effective $M_1-M_2-S_2-Q_2$ system in which we will carry out some of our calculations. This system will be composed of a non-deformable point particle of mass $M_1$ (object $1$) and a deformable body (object $2$) with a mass monopole moment $M_2$, a quadrupole moment $Q_2^{ab}$ and a spin dipole moment $S_2^{a}$; higher-order multipole moments will contribute at higher PN order, so we neglect them here. Given such an effective system, we will then focus on the trajectory of body 2 and calculate how the tidal deformations of body 2 due to the external universe generated by body 1 affect the trajectory of body 2. Once this has been calculated, we will immediately know how the tidal deformations of body 1 due to the external universe generated by body 2 affect the trajectory of body 1 through a symmetry exchange of labels $1 \leftrightarrow 2$ in the tidal terms. With that in hand, we will then be able to compute the equations of motion in the CoM frame for our physical system of two orbiting deformable bodies.   
\subsection{Dynamical equations of motion}\label{subsec:EoM}
The 1PN-accurate equations of motion for objects with arbitrary multipole moments was first derived by DSX assuming that the 1 PN approximation was valid throughout the spacetime~\cite{DSX-I,DSX-II,DSX-III}.
Racine and Flanagan then extended this approach to objects with arbitrarily strong internal structure moving in a PN environment in~\cite{Racine_2005}.
VF provided a simplified set of equations for the $M_1-M_2-S_2-Q_2$ system in the CoM frame and  showed how the equations of motion can be derived from a generalized Lagrangian. We now summarize their method here.

We denote the relative acceleration in the CoM frame by $a^i = a^{i}_{2} - a^{i}_{1}$, the relative velocity by $v^i = v^i_{2} - v^i_{1}$, and the relative normal vector by $n^i = n^i_{2} - n^i_{1}$, where $a^{i}_{A}$, $v^{i}_{A}$ and $n^{i}_{A}$ are the acceleration, the velocity and the normal vector of body $A$. The CoM acceleration of the effective $M_1-M_2-S_2-Q_2$ system can be decomposed into 3 different pieces
\begin{align}\label{eq:aiQ-schematic}
    a^i = a^i_{M} + a^i_{S_{2}} + a^i_{Q_{2}}\,,
\end{align}
where $a^i_M$ is the monopolar (point-particle) contribution (including the mass monopole pieces of both bodies)
\begin{align}\label{eq:aiM-eqn}
    a^i_M 
    &= -\frac{M}{r^2} n^i -
    \frac{1}{c^2}\frac{M}{r^2}
    \bigg\{ 
    n^i \bigg[(1+3\eta) v^2
    -
    \frac{3\eta}{2} \dot{r}^2
    \nonumber\\
    &-
    2(2+\eta) \frac{M}{r}
    \bigg]
    -
    2(2-\eta) \dot{r} v^i
    \bigg\}
    +
    \mathcal{O}\left(c^{-4}\right)
    \,,
\end{align}
$a^i_{S_{2}}$ denotes the contribution due to the coupling between the orbital angular momentum and the spin dipole moment of object $2$ (i.e.~a spin-orbit coupling)
\begin{align}\label{eq:aiS-eqn}
    a^i_{S_{2}}
    &=
    \frac{\epsilon_{abc} S_2^c}{c^2 X_2 r^3}
    \left[ 
    (3+X_2) v^a \delta^{bi}
    -
    3(1+X_2) \dot{r}n^a \delta^{bi}
    +
    6 n^{ai} v^b
    \right]
    \nonumber\\
    &+
    \mathcal{O}\left(c^{-4}\right)\,,
\end{align}
and $a^i_{Q_{2}}$ denotes the contribution from the quadrupolar interaction of object 2
\begin{widetext}
\begin{align}\label{eq:aiQ-eqn}
    a^i_{Q_{2}} &= - \frac{3Q_{2,ab}}{2X_2 r^4}\left[5n^{abi}-2 n^a \delta^{bi} \right]
+\frac{1}{c^2}\Bigg\{ \frac{Q_{2,ab}}{r^4} \bigg[ n^{abi} \left( B_1 v^2 + B_2 \dot{r}^2 + B_3 \frac{M}{r} \right)
\nonumber\\
&+ n^a \delta^{bi} \left( B_4 v^2 + B_5 \dot{r}^2 + B_6 \frac{M}{r} \right)
+ B_7 \dot{r} n^{ab} v^i + B_8 n^a v^{bi} + B_9 \dot{r} n^{ai} v^b + B_{10} v^{ab} n^i + B_{11} \dot{r} v^a \delta^{bi} \bigg]  
\nonumber\\
&+ \frac{\dot{Q}_{2,ab}}{r^3} \left[ B_{12} n^{ab}v^i + B_{13} \dot{r} n^{abi} + B_{14} n^{ai} v^b 
+ B_{15} v^a \delta^{bi} + B_{16} \dot{r} n^a \delta^{bi} \right]
+ \frac{\ddot{Q}_{2,ab}}{r^2}\left[B_{17} n^{abi} + B_{18} n^a \delta^{bi} \right]\Bigg\} +O(c^{-4})
\,.
\end{align}
\end{widetext}
The coefficients $B_{i}$ depend on the component masses and are provided explicitly in Appendix~\ref{appendix:Bi-coeffs}.
Observe that we do not include the spin-spin terms or terms that couple the spin and the quadrupolar interaction because, as mentioned earlier, they would contribute at higher PN order.

In the absence of tidal interactions, the mass and spin of object 2 is conserved, but in their presence, tidal interactions extract energy from the orbit and lead to the evolution of the mass and spin of the object.
In the effective $M_1-M_2-S_2-Q_2$ system, the equations governing this interaction are
\begin{subequations}
\begin{align}
\label{eq:tidal-evolution-spin-mass}
    \partial_t S_{2}^i &= \epsilon^{i}{}_{jk} Q_2^{ja} G_{2,a}^{k} + \mathcal{O}\left(c^{-2}\right)\,,\\
    \label{eq:tidal-evolution-mass}
    \partial_t M_2 &= -\frac{1}{c^2} \left( 
    G_2^{ij}\partial_t Q_{2,ij}
    +
    \frac{3}{2} Q_{2,ij} \partial_t G_2^{ij}\right)
    +
    \mathcal{O}\left(c^{-3}\right)\,,
\end{align}
\end{subequations}
where $G_2^{ij}$ is the  tidal field tensor experienced by body 2 due to body 1.
The 1PN accurate expression for the tidal field tensor is
\begin{align}\label{eq:DSX-1PN-tidal-moment}
    G_2^{ab}
    &=
    \frac{3 M_1}{r^3} n^{<ab>}
    \nonumber\\
    &+
    \frac{3 M_1 }{r^3c^2}
    \bigg[ 
    \bigg( 2 v^2 - \frac{5 X_2^2}{2} \dot{r}^2
    -
    \frac{5 + X_1}{2} \frac{M}{r}
    \bigg) n^{<ab>}
    \nonumber\\
    &+
    v^{<ab>}
    -
    (3-X_2^2) \dot{r} n^{<a}v^{b>}
    \bigg]
    +
    \mathcal{O}\left(c^{-3}\right)\,.
\end{align}

Before we proceed, note that Eqs.~\eqref{eq:aiM-eqn}-\eqref{eq:aiQ-eqn} differs from Eq.(5.9 d) of VF.
The reason is because VF chose to split the mass $M_2$ as
\begin{align}\label{eq:M2-split-VF}
    M_2 &= {}^{n}{}M_2 + \frac{1}{c^2}\left( E_{2,\mathrm{int}} + 3 U_{Q_{2}}\right) + \mathcal{O}\left(c^{-3}\right)\,,
\end{align}
where the ``Newtonian'' mass ${}^{n}{}M_2$ is conserved $\partial_t\left({}^{n}{}M_2\right) =0$.
The ``internal energy'' $E_{2,\mathrm{int}}$ and the tidal potential energy contributions are given by
\begin{subequations}
\begin{align}
    \partial_t \left(E_{2,\mathrm{int}}\right)
    &=
    \frac{1}{2} G_{2,ij} \partial_t Q^{ij}_2\,,\\
\label{eq:UQ-eqn}
    U_{Q_{2}}(t) &= 
    -\frac{1}{2} G_{2,ij} Q^{ij}_2 \,.
\end{align}
\end{subequations}
We do not make this distinction when writing down Eq.~\eqref{eq:aiQ-eqn}, and instead, we account for the evolution of $M_2$ due to tidal interaction directly in the equations of motion.

With this in hand, one can now easily return to our physical system of two deformable bodies. The tidal corrections to the acceleration of body 1 due to the quadrupolar deformations it experiences due to the tidal field of body 2 are simply 
\begin{equation}
a_{S_{1}} = \hat{{\cal{E}}}\left[ a_{S_{2}} \right]\,, \qquad
a_{Q_{1}} = \hat{{\cal{E}}}\left[ a_{Q_{2}} \right]\,,
\end{equation}
where we have defined the label exchange operator $\hat{{\cal{E}}}\left[\cdot\right]$ as one which makes the transformation $1 \leftrightarrow 2$ in the labels of its argument. For example, $\hat{{\cal{E}}}\left[X_{2}\right] = X_{1}$ and $\hat{{\cal{E}}}\left[q\right] = 1/q$ . Notice that when acting the label exchange operator on relative 3-vector quantities, like $v^{i}$ and $n^{i}$, on incurs a minus sign due to their definitions. The total relative acceleration of the binary system in the CoM frame is then simply
\begin{align}\label{eq:aiQ-schematic-final}
    a^i = a^i_{M} + a^i_{S_{1}}  + a^i_{S_{2}} + a^i_{Q_{1}} + a^i_{Q_{2}}\,.
\end{align}
The evolution of the mass monopole $M_{1}$ and the spin dipole $S_{1}$, which implicitly appear in the above equation, can be obtained by acting the label exchange operator on Eqs.~\eqref{eq:tidal-evolution-spin-mass} and~\eqref{eq:tidal-evolution-mass}. 

\subsection{Internal dynamics and tidal response}\label{sec:tidal-response}
To close the dynamical equations [Eqs.~\eqref{eq:aiQ-schematic}, \eqref{eq:tidal-evolution-spin-mass} and~\eqref{eq:tidal-evolution-mass}] in the effective $M_1-M_2-S_2-Q_2$ system, we need to describe the dynamics of the quadrupole moment $Q_2^{ab}$.
Linear response theory can be used to model the tidal response in the body frame of the object (see Sec. III of~\cite{HegadeKR:2024agt} for more details) as
\begin{align}
    Q_2^{ab} \left(s_2(t)\right)
    \! \propto \!
    \int_{-\infty}^{\infty} \!\!\!\!
    K_2\!\left(s_2(t)- s_2(t')\right) \;
    G_2^{ab}\!(s_2(t'))
    \left(\frac{dt'}{ds_2}\right) ds_2
    \,,
\end{align}
where $K_2\!\left(\cdot\right)$ is the tidal response function of the object and $G_2^{ab}\!\left(\cdot\right)$ is the tidal moment of the object [Eq.~\eqref{eq:DSX-1PN-tidal-moment}].
The function $s_2(t)$ denotes the dependence of the local time in the body frame as a function of the global time coordinate. The factor $(dt/ds_2)$ is then the local redshift factor and one can show that
\begin{align}
    \left(\frac{dt}{d s_2}\right)
    &=1 + \frac{v^2 X_1^2}{c^2} + \frac{M_1}{r c^2} 
    +
    \mathcal{O}\left(c^{-4}\right)
    \,,
\end{align}
by using the 1PN accurate coordinate transformations in Eq.(2.17) of~\cite{Racine_2005}, and specializing these coordinate transformations to the body adapted gauge of VF (see the discussion above Eq.(3.45) of~\cite{Racine_2005}). Since we are interested in the inspiral, we can assume weak tidal interactions and truncate the tidal response function in a small frequency approximation. 
We retain only the leading order conservative and dissipative evolution for the quadrupole moment and ignore contributions from the spin of the object to obtain
\begin{align}\label{eq:Q2-diss-eqn}
    Q_2^{ab} &= 
    \frac{\Lambda_2 X_2^5 M^5}{c^{10}} G_{2}^{ab}\left(1+\mathcal{O}\left(c^{-3}\right)\right)
    \nonumber\\
    &
    -\frac{\Xi_2}{c^{13}} X_2^6 M^6 \left(\frac{dt}{d s_2} \right)\partial_t G_2^{ab} \left(1+\mathcal{O}\left(c^{-3}\right)\right)
    \,,
\end{align}
where $\Lambda_2$ is the conservative tidal deformability of object 2, and $\Xi_2$ is the dissipative tidal deformability of object 2. We refer to these terms as \emph{conservative} and \emph{dissipative} because they are even and odd under time reversal.
We have uncontrolled remainders of $\mathcal{O}(c^{-3})$ in the expression above because the tidal field [Eq.~\eqref{eq:DSX-1PN-tidal-moment}] has uncontrolled remainders of $\mathcal{O}(c^{-3})$.
Returning to the physical problem, a similar expression holds for the quadrupole moment tensor of body 1 by acting the label exchange operator on Eq.~\eqref{eq:Q2-diss-eqn}.

The value of the conservative tidal deformability $\Lambda_2$ depends only on the equation of state of the object and has been extensively studied in the literature~\cite{Chatziioannou_2020}.
The value of the dissipative tidal deformability depends not only on the equation of state, but also on the details of the internal dissipative mechanism of the object.
For example, if object 2 is a black hole, then the internal dissipative mechanism is the absorption of gravitational waves by the black hole's event horizon;
the value of $\Xi_2$ for a rotating and non-rotating black holes can be found in~\cite{Poisson_2004,Saketh:2023bul}.
On the other hand, if the object is composed of a viscous fluid, then the value of $\Xi_2$ depends on the shear and bulk viscosity of the fluid.
A method for calculating the value of $\Xi_2$ for a relativistic viscous fluid object was described recently by us in~\cite{HegadeKR:2024agt}, which greatly extends the previous Newtonian work of~\cite{Lai:1993di}. As explained in~\cite{HegadeKR:2024agt}, $\Xi_{2}$ can be obtained by solving the perturbed Einstein equations, once an equation of state and a viscosity profile have been prescribed; the latter requires the modeling of internal nuclear dissipative rates, such as Urca reactions~\cite{Jones:2001ya,Arras:2018fxj,Lindblom:2001hd,Gusakov:2008hv,Alford:2020pld,Most:2021zvc,Yang:2023ogo,Chabanov:2023abq,chabanov2023impact}.
This is an involved calculation that is currently being tackled separately~\cite{future-work-viscosity-nuclear}. Therefore, in this paper, we will not concern ourselves with the mapping between $\Xi_{1,2}$ and the shear or bulk viscosities of the stars, deferring such an analysis to future work.  

We introduce the following notation for the conservative and dissipative sectors of the quadrupole moment
\begin{subequations}
\begin{align}
    Q_2^{\mathrm{cons},ab} &= \frac{\Lambda_2 X_2^5 M^5}{c^{10}} G_{2}^{ab} \,,\\
    Q_2^{\mathrm{diss},ab} &= -\frac{\Xi_2}{c^{13}} X_2^6 M^6 \left(\frac{dt}{d s_2} \right)\partial_t G_2^{ab}
    \,,
\end{align}
\end{subequations}
and similar expressions for the quadrupole deformations of body $1$, obtained by applying the label exchange operator to the above equations. These expressions will be used to simplify some equations below.
\subsection{Generalized Lagrangian and energy balance relationship}\label{sec:generalized-lagrangian}
To evaluate the gravitational wave phase and flux, we need to derive an energy-balance relation from the orbital equation of motion.
To do this, we first derive the equations of motion [Eq.~\eqref{eq:aiQ-schematic}] from a generalized \textit{acceleration dependent} Lagrangian.
The Lagrangian $\mathcal{L}$ of the effective $M_1-M_2-S_2-Q_2$ system can be split into three contributions~\cite{Vines_flanagan_2010} 
\begin{align}\label{eq:generalized-lagrangian}
    \mathcal{L} &= \mathcal{L}_{M} + \mathcal{L}_{S_{2}} + \mathcal{L}_{Q_{2}}\,,
\end{align}
where the monopolar piece is given by
\begin{align}\label{eq:LM-eqn}
    \mathcal{L}_{M} &= 
    \frac{\mu M v^2}{2}
    +
    \frac{\mu M}{r}
    +
    \frac{\mu}{c^2}
    \bigg\{ 
    \frac{1-3\eta}{8} v^4
    \nonumber\\
    &+
    \frac{M}{2r}
    \bigg[ 
    (3+\eta) v^2
    +
    \eta \dot{r}^2
    -
    \frac{M}{r}
    \bigg]
    \bigg\}
    +
    \mathcal{O}\left(c^{-4}\right)
    \,.
\end{align}
the spin contribution due to body $2$ is given by 
\begin{align}\label{eq:LS-eqn}
    \mathcal{L}_{S_{2}} &=
    \frac{X_1 \epsilon_{abc} S_2^a v^b}{c^2}
    \left( 
    2\frac{M}{r} n^c
    +
    \frac{X_1}{2} a^c
    \right)
    +
    \mathcal{O}\left(c^{-4}\right)\,,
\end{align} 
and the quadrupolar contribution due to body $2$ is
\begin{align}\label{eq:LQ-eqn}
    \mathcal{L}_{Q_{2}} &= 
    \frac{3 M_1 Q_2^{ab} n^a n^b}{2r^3}
    +
    \frac{1}{c^2}
    \bigg\{ 
    \frac{M Q_2^{ab}}{r^3}
    \bigg[ 
    n^{a} n^{b}
    \nonumber \\
    & \times \bigg( 
    A_1 v^2 + A_2 \dot{r}^2 + A_3 \frac{M}{r}
    \bigg)
    +
    A_4 v^{a} v^{b}
    +
    A_5 \dot{r} n^a v^b
    \bigg]
    \nonumber\\
    &+
    \frac{M \dot{Q}_2^{ab}}{r^2}
    \bigg[ 
    A_6 n^a v^b + A_7 \dot{r} n^a n^b
    \bigg]
    \nonumber \\
    &-
    3 U_{Q_{2}}
    \bigg[ 
    A_8 v^2 + A_9 \frac{M}{r}
    \bigg]
    \bigg\}
    +
    \mathcal{O}\left(c^{-4}\right)
    \,.
\end{align}
Note that the tidal potential energy due to the quadrupole moment $U_{Q_{2}}$ [see Eq.~\eqref{eq:UQ-eqn}] is to be understood as a function of time in the above expression, i.e.,
the expression for $G_{2,ab}$ [Eq.~\eqref{eq:DSX-1PN-tidal-moment}] in Eq.~\eqref{eq:UQ-eqn} is only substituted in after the Lagrangian is derived and the equations of motion are obtained. 
The same comments apply to the quadrupole moment $Q_2^{ab}(t)$.
The physical reason behind this is that the tidal moment $G_2^{ab}(t)$ are obtained by evaluating on the worldlines of the objects in a matched asymptotic expansion.
We first write down the field equations and then obtain the tidal moments by matching the gravitational metric to the external field in the reference frame of the object.
We are therefore interested in the dynamics of the objects after integrating out the gravitational field.
The technical term for this procedure is the method of reduced actions.
We refer the reader to Sec. II E of~\cite{Bini-Damour-Faye} for a summary of this technique.
In practice, we have obtained the Lagrangians displayed above by writing down all possible combinations of the terms that appear at $1$ PN order and then matching the equations of motion derived from the Lagrangian to Eq.~\eqref{eq:aiQ-schematic}.

The generalized Lagrangian approach to deriving the quadrupolar interaction was first followed by VF. Our approach differs nominally from their approach because we do not split the mass of object 2 as given in Eq.~\eqref{eq:M2-split-VF}.
This difference then leads to the appearance of $U_{Q_{2}}(t)$ in Eq.~\eqref{eq:LQ-eqn} ,while Eq.(5.10 d) of VF has a term proportional to $E_{2,\mathrm{int}}$ in the Lagrangian. The coefficients $A_i$ are the same in both our expressions and those of VF, and thus, we list them in Appendix~\ref{appendix:Ai-coeffs}.

The equations of motion can be derived from the generalized Lagrangian by using the generalized Euler-Lagrange equation
\begin{align}
    -\frac{d^2}{dt^2} \left(\frac{\partial\mathcal{L}}{\partial a^i}\right)
    +
    \frac{d}{dt} \left(\frac{\partial\mathcal{L}}{\partial v^i}\right)
    -
    \frac{\partial\mathcal{L}}{\partial z^i}
    &=0\,.
\end{align}
For solutions to the equations of motion, a generalized energy balance law can also be obtained 
\begin{align}\label{eq:generalized-energy-v1}
    \frac{d}{dt} \mathcal{E}
    &=
    -
    \frac{\partial \mathcal{L}}{\partial t}
    \,,
\end{align}
where the generalized energy is a function of the generalized Lagrangian evaluated on shell
\begin{align}
    \mathcal{E}\left[ \mathcal{L}\right]
    &=
    a^i \frac{\partial\mathcal{L}}{\partial a^i}
    -
    v^i \frac{d}{dt} \frac{\partial\mathcal{L}}{\partial a^i}
    +
    v^i \frac{\partial\mathcal{L}}{\partial v^i}
    -
    \mathcal{L}
    \,.
\end{align}
This generalized energy is not to be confused with the orbital energy, and instead, it should be thought more of as a ``Hamiltonian''. 

We can obtain orbital energy-balance law from the above generalized energy balance law if we implement a separation of conservative and dissipative terms. As we mentioned before, by ``conservative'' and ``dissipative'' we mean terms that are even and odd under time reversal. Therefore, it is natural to decompose the quadrupolar part of the Lagrangian into conservative and dissipative contributions 
\begin{align}
\label{eq:LQ2}
    \mathcal{L}_{Q_{2}} &= \mathcal{L}_{Q_{2},\mathrm{cons}} + \mathcal{L}_{Q_{2},\mathrm{diss}}\,,
\end{align}
where
\begin{subequations}
\begin{align}
    \mathcal{L}_{Q_{2},\mathrm{cons}}
    &=
    \mathcal{L}_{Q_{2}}\left[Q^{{\rm{cons}},ab}_2\right]\,,
    \\
    \mathcal{L}_{Q_{2},\mathrm{diss}}
    &=
    \mathcal{L}_{Q_{2}}\left[Q^{{\rm{diss}},ab}_2\right]\,,
\end{align}
\end{subequations}
and where $Q^{{\rm{cons}},ab}_2$ and $Q^{{\rm{diss}},ab}_2$ are the terms proportional to $\Lambda_{2}$ and the $\Xi_{2}$ in Eq.~\eqref{eq:Q2-diss-eqn} respectively. Similar decompositions can be applied in the physical system to find the quadrupolar part of the Lagrangian of body 1, ${\cal{L}}_{Q_{1}}$ simply by acting the exchange label operator on Eq.~\eqref{eq:LQ2}. With this decomposition, Eq.~\eqref{eq:generalized-energy-v1} becomes
\begin{align}\label{eq:generalized-energy-v2}
     \frac{dE_{\mathrm{orb}}}{dt} := \frac{d}{dt} \mathcal{E}
    + 
    \frac{\partial \mathcal{L}_{\rm cons}}{\partial t} &=
    -
    \frac{\partial \mathcal{L}_{\rm diss}}{\partial t} := \mathcal{F}_{\mathrm{diss}}
    \,,
\end{align}
where we have identified the left-hand side as the rate of change of the orbital energy with respect to time and the right-hand side with a dissipative flux.


Let us tackle the left- and right-hand sides of this energy-balance equation separately. The orbital energy can be written as
\begin{align}\label{eq:Eorb-eqn}
    E_{\mathrm{orb}}
    &=
    \mathcal{E}\left[ \mathcal{L}_M + \mathcal{L}_{S_{1}} + \mathcal{L}_{Q_{{1}},\mathrm{cons}} +  \mathcal{L}_{S_{2}} + \mathcal{L}_{Q_{{2}},\mathrm{cons}} \right]
    \nonumber \\
    &+
    \int \left(\frac{\partial \mathcal{L}_{Q_{{1}},\mathrm{cons}}}{\partial t}  + \frac{\partial \mathcal{L}_{Q_{{2}},\mathrm{cons}}}{\partial t} \right)dt
    \nonumber\\
    & =  E_{\mathrm{orb,M}} + E_{\mathrm{orb,S}} + E_{\mathrm{orb,Q}}\,,
\end{align}
where we have decomposed the orbital energy into a piece that depends on the monopole part of the Lagrangian, another one that depends on the spin dipole moment, and a third that depends on the conservative part of the quadrupole moment. Evaluating each of these pieces, one finds
\begin{align}
    E_{\mathrm{orb,M}}  &=
    \frac{\mu v^2}{2}
    -
    \frac{M \mu}{r}
    +
    \frac{\mu}{c^2}
    \bigg\{ 
    \frac{3}{8}(1-3\eta) v^4
    \nonumber \\
    &+
    \frac{M}{2r}
    \bigg[
    (3+\eta) v^2
    +
    \eta \dot{r}^2
    +
    \frac{M}{r}
    \bigg]
    +
    \mathcal{O}\left(c^{-2}\right)
    \bigg\}\,,
     \\
    E_{\mathrm{orb,S}} &= 
    -
    \frac{1}{c^2}
    \left[
    \frac{X_1^2 M }{r^2} \epsilon_{abc}n^a S_2^b v^c
    +
    \mathcal{O}\left(c^{-2}\right)
    \right] + 1 \leftrightarrow 2\,,
    \\
    E_{\mathrm{orb,Q}}  &=
    -\frac{3 M^7 X_1^2 X_2^5 \Lambda_2}{2 r^6 c^{10}}
    \bigg\{ 
    1
    +
    \frac{1}{c^2}
    \bigg[ 
    -\frac{2 M}{r}
    \nonumber \\
    &+3 \dot{r}^2 \left(X_1^2+4 X_1-4\right)-v^2 X_1 (2 X_1+3)
    \bigg]
    \nonumber \\
    &+
    \mathcal{O}(c^{-3})
    \bigg\} + 1 \leftrightarrow 2
    \,.
\end{align}
Not surprisingly, the orbital energy [Eq.~\eqref{eq:Eorb-eqn}] is the same as the 1PN accurate conserved orbital energy for purely conservative tidal interactions 
(see\footnote{The same comments made above Eq.~\eqref{eq:M2-split-VF} apply when comparing  Eq. (2.7) of~\cite{Vines_Waveform_2011} to Eq.~\eqref{eq:Eorb-eqn}} Eq.~(2.7) of~\cite{Vines_Waveform_2011}).
However, note that unlike the case of purely conservative tidal interaction, the mass monopole moment $M_2$ and the spin $S_2$ are \textit{not} conserved [see Eqs.~\eqref{eq:tidal-evolution-spin-mass} and ~\eqref{eq:tidal-evolution-mass} for their evolution equations].

The flux due to internal dissipation can be further decomposed into
\begin{align}
 \mathcal{F}_{\mathrm{diss}}
    =
     -
    \frac{\partial \mathcal{L}_{\rm diss}}{\partial t}
    =
    \mathcal{F}_{\mathrm{diss,M}}
    +
    \mathcal{F}_{\mathrm{diss,S}}
    +
    \mathcal{F}_{\mathrm{diss,Q}}\,
\end{align}
to separate the part of the flux that comes from the monopole term, from that which comes from the spin dipole moment and the quadrupole moment. More specifically, the dissipative mass flux $\mathcal{F}_{\mathrm{diss,M}}$ is given by 
\begin{align}
    \mathcal{F}_{\mathrm{diss,M}}
    &=
    -
    \frac{\partial \mathcal{L}_M}{\partial t} 
    \,,\nonumber\\
    &=
    \frac{\partial_t{(\mu M)}}{2} v^2
    +
    \frac{\partial_t(\mu M)}{r}
    +
    \mathcal{O}\left(c^{-4}\right)
    \,,
\end{align}
the dissipative spin flux $\mathcal{F}_{\mathrm{diss,S}}$ is given by
\begin{align}
    \mathcal{F}_{\mathrm{diss,S}}
    &=
    -
    \frac{\partial \mathcal{L}_{S_{1}}}{\partial t}  -
    \frac{\partial \mathcal{L}_{S_{2}}}{\partial t}
    \,, \nonumber\\
    &=
    \frac{3 M^2X_1^2(4-X_1) }{2 c^2 r^5}
    n^b Q_2^{\mathrm{diss},bc} \left(n^c \dot{r} - v^c\right)
    \nonumber\\
    &+
    \mathcal{O}\left(c^{-4}\right) + 1 \leftrightarrow 2
    \,,
\end{align}
and the dissipative quadrupolar flux $\mathcal{F}_{\mathrm{diss,Q}}$ is given by
\begin{widetext}
\begin{align}\label{eq:FdissQ-eqn}
    \mathcal{F}_{\mathrm{diss,Q}}
    &=
    -\frac{d}{dt} \left(\mathcal{E}\left[\mathcal{L}_{Q_{1},\mathrm{diss}}\right]\right)
    -
    \frac{\partial \mathcal{L}_{Q_{1},\mathrm{diss}}}{\partial t}
     -\frac{d}{dt} \left(\mathcal{E}\left[\mathcal{L}_{Q_{1},\mathrm{diss}}\right]\right)
    -
    \frac{\partial \mathcal{L}_{Q_{2},\mathrm{diss}}}{\partial t}
    \,, \nonumber\\
    &=
    -\frac{3M X_1}{2 r^4} n^a Q_2^{{\rm{diss}},ab} \left(5 n^b \dot{r}-2 v^b\right)
    \nonumber\\
    &+
    \frac{1}{c^2}
    \bigg\{ 
    \frac{Q_2^{{\rm{diss}},ab} M }{r^4}
    \bigg[ 
    n^{a}n^{b} \dot{r}
    \bigg(
    \alpha_1 v^2 
    +
    \alpha_2 \dot{r}^2
    +
    \alpha_3 \frac{M}{r}
    \bigg)
    +
    \dot{r} \alpha_4 v^a v^b 
    +
    n^a v^b
    \bigg(
    \alpha_5 v^2 
    +
    \alpha_6 \dot{r}^2
    +
    \alpha_7 \frac{M}{r}
    \bigg)
    \bigg]
    \nonumber\\
    &+
    \frac{\dot{Q}_2^{{\rm{diss}},ab} M }{r^3}
    \bigg[
    n^a n^b
    \bigg( 
    \alpha_8 v^2 + \alpha_9 \dot{r}^2
    \bigg)
    +
    \alpha_{10} v^a v^b
    +
    \alpha_{11} \dot{r} n^a v^b
    \bigg]
    +
    \frac{\ddot{Q}_2^{{\rm{diss}},ab} M }{r^2}
    \bigg[ 
    \alpha_{12} \dot{r} n^a n^b  
    +
    \alpha_{13} n^a v^b 
    \bigg]
    \bigg\}
    +
    \mathcal{O}\left(c^{-4}\right)  
    \nonumber \\
    & + 1 \leftrightarrow 2\,.
\end{align}
\end{widetext}
where an overhead dot denotes derivatives with respect to the global time coordinate $t$, and the coefficients $\alpha_i$ are listed in Appendix~\ref{appendix:alphai-coeffs}.
The total dissipative flux can be further simplified by substituting Eq.~\eqref{eq:Q2-diss-eqn} to obtain
\begin{widetext}
\begin{align}\label{eq:viscous-flux-1PN}
     \mathcal{F}_{\mathrm{diss}}
     &=
     - \frac{9 \Xi_{2}{} M^8  X_1^2 X_2^6}{c^{13} r^8}
     \bigg\{(2 \dot{r}^2 + v^2)
     +
     \frac{1}{c^2}
     \bigg[
     \frac{2 M^2 X_1}{r^2} + \frac{M \dot{r}^2 (-86 + 94 X_1 + X_1^2)}{2 r} + v^4 (-2 + 7 X_1 + \tfrac{3}{2} X_1^2) 
     \nonumber\\
     &+ 5 \dot{r}^4 (-25 + 28 X_1 + 2 X_1^2) + v^2 \bigl(\dot{r}^2 (76 - 108 X_1 - 5 X_1^2) -  \frac{M (2 + 22 X_1 + X_1^2)}{2 r}\bigr)
     \bigg] 
     \bigg\} + 1 \leftrightarrow 2
     \,.
\end{align}
\end{widetext}
The leading PN order expression for the dissipative flux [Eq.~\eqref{eq:viscous-flux-1PN}] was derived by us in Eq. (24b) of~\cite{Ripley_2023} and it is the familiar Newtonian tidal heating term. The 1PN correction to this leading PN order expression is new and derived here for the first time. We also write down the simplified expressions for the spin and mass evolution equations [Eqs.~\eqref{eq:tidal-evolution-spin-mass} and~\eqref{eq:tidal-evolution-mass}] after substituting Eq.~\eqref{eq:Q2-diss-eqn}
\begin{subequations}\label{eq:tidal-evolution-spin-mass-subs}
\begin{align}
\label{eq:tidal-evolution-spin-subs}
    \partial_t S^i_2 &=
    \frac{9\Xi_2M^8 X_1^2 X_2^6}{r^7 c^{13}}\epsilon^{i}_{aj} n^a v^j 
    \left[
    1
    +
    \mathcal{O}\left(c^{-2}\right)
    \right]
    \,,\\
    \label{eq:tidal-evolution-mass-subs}
    \partial_t M_2 &= 
    \frac{9 \Xi_2 M^8 X_1^2 X_2^6}{r^9 c^{15}}
    \left( 2M + 18 r \dot{r}^2 - r v^2\right)
    \left[1
    +
    \mathcal{O}\left(c^{-2}\right)
    \right]
    \,,
\end{align}
\end{subequations}
since these equations were used to simplify some of the terms when deriving Eq.~\eqref{eq:viscous-flux-1PN}. Similar expressions can be obtained for the spin and mass evolution equations of body 1 through the action of the label exchange operator on the above equations. 
\section{Gravitational wave phase for a circular orbit}\label{sec:GW-phase}
In this section, we specialize to the case of quas-circular orbits and calculate the gravitational wave phase for the dissipative tidal interaction to 1PN order beyond the leading PN order tidal effects in the phase.
The leading PN order corrections to the phase due to conservative tidal interactions were presented in~\cite{Flanagan:2007ix,Vines_Waveform_2011,Henry_2020_phase}.
We are mainly interested in deriving the gravitational wave flux due to dissipative tidal interactions, so we will \textit{ignore} the conservative tidal contribution from hereon and set $\Lambda_2$ to zero; we will restore $\Lambda_{2}$ at the end of our calculation.  
We also assume that the dynamics of the $M_1-M_2-S_2-Q_2$ system can be specified to leading order using point particle dynamics, which evolves adiabatically under the influence of gravitational wave and tidal dissipation.
Finally, we assume that objects are initially non-spinning.

With these assumptions, we add the gravitational wave dissipation into the energy-balance law [Eq.~\eqref{eq:generalized-energy-v2}], 
\begin{align}\label{eq:energy-balance-law-internal-GW-dissipation}
    \frac{dE_{\mathrm{orb}}}{dt}
    =&
    \mathcal{F}_{\mathrm{diss}}
    +
    \mathcal{F}_{\mathrm{GW}}\,,
\end{align}
where the gravitational wave flux is given by [see e.g.~Eq.~(454) of~\cite{blanchet2024postnewtonian}]
\begin{align}
    \mathcal{F}_{\mathrm{GW}}
    &=
    -\frac{1}{5c^5} \left( \partial_t^3 Q^{ij}_{\mathrm{sys}} \right)^3
    \nonumber\\
    &-
    \frac{1}{c^7}
    \left[\frac{1}{189}  \left( \partial_t^4 Q^{ijk}_{\mathrm{sys}} \right)^3 
    +
    \frac{16}{45} \left( \partial_t^3 S^{ij}_{\mathrm{sys}} \right)^3 
    \right]
    +
    \mathcal{O}\left(c^{-8}\right)\,.
\end{align}
In the above equation, $Q^{ij}_{\mathrm{sys}}$ is the mass quadrupole moment, $Q^{ijk}_{\mathrm{sys}}$ the mass octupole moment and $S^{ij}_{\mathrm{sys}}$ is the current quadrupole moment of the system.
To calculate these quantities, we can ignore the purely dissipative quadrupolar interaction, as it would contribute as a high PN order correction to the orbital energy instead of the flux. The point-particle contributions are well known (see e.g.~Eqs.~(445) and (451) of~\cite{blanchet2024postnewtonian}). 

The equations governing the motion to two point particles moving around each other in a circular orbit of radius $r$ and orbital frequency $F$ is given by
\begin{align}\label{eq:circ-orb-def}
    r&= 
    \frac{M^{1/3}}{\omega^{2/3}} -
    \frac{M(3-\eta)}{3c^2}
    +
    \mathcal{O}(c^{-3})\,,
\end{align}
where $\omega = 2 \pi F$ and the $\mathcal{O}(c^{-3})$ contributions are due to the spin of the object~\cite{blanchet2024postnewtonian}. We are allowed to use this version of Kepler's third law at 1PN order because the dissipative tidal effects only enter through the dissipative flux, and do not modify the orbital energy (unlike the conservative tidal deformations).  
Using the adiabatic evolution assumption, we can rewrite Eq.~\eqref{eq:energy-balance-law-internal-GW-dissipation} as 
\begin{align}\label{eq:energy-balance-averaged}
    & \left<\frac{d E_{\mathrm{orb}}}{dx}\right> \dot{x}
    +
    \left<\frac{d E_{\mathrm{orb}}}{d M_2}\right> \dot{M}_2
    \nonumber \\
    &
    +
    \left<\frac{d E_{\mathrm{orb}}}{d S_2^{b}}\right> \dot{S}_2^{b}
    =
    \left< \mathcal{F}_{\mathrm{GW}}\right>
    +
    \left< \mathcal{F}_{\mathrm{diss}}\right>
    \,,
\end{align}
where
\begin{align}\label{eq:x-def}
    x \equiv \left(\frac{2 \pi M F}{c^3}\right)^{2/3}  = \left(\frac{M\omega}{c^3}\right)^{2/3} 
    \,,
\end{align} 
and the angular brackets are used to denote the fact that the expressions are evaluated for a circular orbit.

Let us now evaluate each of the terms that appear in the averaged energy-balance law of Eq.~\eqref{eq:energy-balance-averaged}. 
For circular orbits, the gravitational wave flux to 1PN order can be evaluated to (see, e.g., Eq. (480) of~\cite{blanchet2024postnewtonian})
\begin{align}
    \left< \mathcal{F}_{\mathrm{GW}}\right>
    &=
    -\frac{32 c^5 \eta^2 x^5}{5}
    \bigg[ 
    1
    +
    \left(-\frac{1247}{336} - \frac{35}{12} \eta \right)
    x
    \nonumber\\
    &+
    \mathcal{O}\left(c^{-3}\right)
    \bigg]
    \,.
\end{align}
The 1PN accurate expression for the dissipative flux $\left< \mathcal{F}_{\mathrm{diss}}\right>$, the spin evolution $\dot{S}_2^b$ and the mass evolution $\dot{M}_2$ can be obtained by substituting Eq.~\eqref{eq:circ-orb-def} into Eq.~\eqref{eq:viscous-flux-1PN}, \eqref{eq:tidal-evolution-mass-subs} and \eqref{eq:tidal-evolution-spin-subs} respectively to obtain 
\begin{subequations}\label{eq:flux-all-circ}
\begin{align}
&\left< \mathcal{F}_{\mathrm{diss}}\right>
=
-9 \Xi_{2}{} x^9 X_1^2 X_2^6 c^{5} \bigg[
1
+
\nonumber\\
&x\bigl[3 + X_1^2 - 2 X_1 (1 + X_2)\bigr]
+ 1 \leftrightarrow 2 + 
\mathcal{O}(c^{-4})
\bigg] 
\,,\\
&\left <\dot{M}_2\right>
=
9 \, \Xi_{2}{} x^9 c^{3} X_1^2 X_2^6
\left[1+
\mathcal{O}(c^{-2})\right] \,,\\
&\left<\dot{S}_2^b\right>
=
9 \,\Xi_{2}{} x^7 c\, \epsilon^{b}{}_{ac} M n^{a} v^{c} X_1^2 X_2^6
\left[1+
\mathcal{O}(c^{-2})\right]
\,.
\end{align}
\end{subequations}
To evaluate Eq.~\eqref{eq:energy-balance-averaged} we need to calculate the product of the derivative of the orbital energy with respect to the dipole and the quadrupole moments contracted onto themselves. Doing so, one finds 
\begin{subequations}\label{eq:derivatives-E-orb}
\begin{align}
    &\left<\frac{d E_{\mathrm{orb}}}{d M_2}\right> \left<\dot{M}_2\right>
    =
    \frac{9 c^5(X_1-2) X_1^3 X_2^6 \Xi_2 x^{10}}{2 }
    \bigg[1 \nonumber\\
    &\hspace{3cm}
    + \mathcal{O}(c^{-2})\bigg]
    \,,\\
    &\left<\frac{d E_{\mathrm{orb}}}{d S_2^{b}}\right> \left<\dot{S}_2^b\right>
    =
    9 c^5 X_1^4 X_2^6 \Xi_2 x^{10}
    \left[1+ \mathcal{O}(c^{-2})\right]
    \,.
\end{align}
\end{subequations}
Using that the rate of change of the orbital energy with respect to $x$ is simply 
${d E_{\mathrm{orb}}}/{dx} = -{M X_1 X_2 c^2}/{2 } [1 -{x (9+\eta)}/{6}]$ to 1PN order, and using Eqs.~\eqref{eq:flux-all-circ} and \eqref{eq:derivatives-E-orb} we can simplify Eq.~\eqref{eq:energy-balance-averaged} to obtain
\begin{align}\label{eq:xdot-eqn-circ}
    \dot{x}
    &=
    \left(\left<\frac{d E_{\mathrm{orb}}}{dx}\right>\right)^{-1}
    \bigg\{
    \left< \mathcal{F}_{\mathrm{GW}}\right>
    +
    \left< \mathcal{F}_{\mathrm{diss}}\right>
    \nonumber\\
    &\hspace{1cm}
    -
    \left<\frac{d E_{\mathrm{orb}}}{d M_2}\right> \left<\dot{M}_2\right>
    -
    \left<\frac{d E_{\mathrm{orb}}}{d S_2^{b}}\right> \left<\dot{S}_2^{b}\right>
    \bigg\}
    \,, 
\end{align}
or explicitly,
\begin{align}
&\dot{x} =   \frac{64 x^5 \eta c^3}{5 M}
    \bigg\{ 
    1
    -
    \frac{x}{336}
    \left(743 + 924 \eta\right)
    +
    \mathcal{O}\left(c^{-3}\right)
    \bigg\}  
    \nonumber \\
    &+
    \frac{18 \Xi_2 x^9 X_1 X_2^5 c^3}{M}
    \bigg\{ 
    1
    +
    \frac{x}{6} \left(27+15X_1^2 - X_1(18+11 X_2) \right) 
    \nonumber\\
    &
    \hspace{3cm}
    +
    \mathcal{O}\left(c^{-3}\right)
    \bigg\} 
    \,.
\end{align}
We see that the dissipative tidal contribution goes as $x^9$, so it is a 4PN contribution relative to the leading PN order point-particle contribution, which is proportional to $x^5$. The 1PN corrections on the right-hand side goes as $x^{10}$, so it is a 5PN contribution and it enters at the same order as the conservative tidal effects (which for simplicity we have not written out explicitly in any of the above equations).  

We can use Eq.~\eqref{eq:xdot-eqn-circ} to find the gravitational wave phase in the stationary phase approximation.
We will assume here that both objects are deformable.
We denote the gravitational wave frequency by $f$ and the orbital frequency by $F$.
In the stationary phase approximation, we have that $f=2 F$ (we consider an $\ell=2$ mode) and the gravitational wave Fourier phase $\Psi(f)$, can be written as~\cite{Yunes:2009yz}
\begin{align}
    \label{eq:d2Psi-domega2-original}
    \Psi(f) = 2 \pi f t(f) - 2\phi(f)
\end{align}
where 
\begin{subequations}
\begin{align}
    t(f)
    &=
    t_c + \int^{F=\frac{f}{2}} \frac{1}{\dot{x}} dx
    \,,\\
    \phi(f)
    &=
    \frac{\varphi_c}{2} + \frac{\pi}{8}
    +
    2\pi \int^{F = \frac{f}{2}} \frac{c^3 x^{3/2}}{M} \frac{dt}{dx} dx
    \,.
\end{align}
\end{subequations}
Using the above expressions and Eq.~\eqref{eq:xdot-eqn-circ} we find that
\begin{align}
\label{eq:stationary-phase-approximation-viscosity}
    \Psi\left(f\right)
    &=
     2\pi f t_c
    -
    \varphi_c
    -
    \frac{\pi}{4} + \Psi_{\mathrm{pp}}
    +
    \Psi_{\mathrm{cons}}
    +
    \Psi_{\mathrm{diss}}
    \,,
\end{align}
where we have here restored the conservative tidal effects, $u = (2 \pi f M/ c^3)^{1/3}$ is another PN expansion parameter, $\varphi_c$ is the coalescence phase, and $t_c$ is the coalescence time.
The point particle contribution to the phase to 1PN order is given by
\begin{align}
    \Psi_{\mathrm{pp}}
    &= 
    \frac{3}{128 \eta u^5}
    \bigg\{ 
    1
    +
    \bigg[\frac{3715}{756} + \frac{55 \eta}{6}
    \bigg]
    u^2
    +
    \mathcal{O}\left(c^{-3}\right)
    \bigg\}
    \,;
\end{align}
higher order corrections up to 4.5 PN order can be found in Eq.~(484) of~\cite{blanchet2024postnewtonian}.
The conservative tidal contribution to the phase to 1PN order is given by~\cite{Flanagan:2007ix,Vines_Waveform_2011} 
\begin{align}
    &\Psi_{\mathrm{cons}}
    = 
    \frac{3 u^5}{128 \eta}
    \bigg[ 
    \left(-\frac{39}{2} \tilde{\Lambda} \right) 
    \nonumber\\
    &+
    \left( 
    -\frac{3115}{64} \tilde{\Lambda}
    +
    \frac{6595}{364}
    \sqrt{1-4\eta} \delta \tilde{\Lambda}
    \right) u^2
    +
    \mathcal{O}{(c^{-3})}
    \bigg]
    \,,
\end{align}
where $\tilde{\Lambda}$ and $\delta \tilde{\Lambda}$ are effective combinations of the conservative tidal deformabilities of each of the stars (see Eqs.(5) and (6) of~\cite{Wade_2014}).
higher order corrections up to 7PN order can be found in~\cite{Henry_2020_phase}. 
The dissipative tidal contribution to the phase is given by
\begin{align}
    \Psi_{\mathrm{diss}}
    &=
    \frac{3 u^3}{128 \eta}
    \bigg\{ 
    \frac{25}{4}\Xi_2 X_2^4
    -\frac{75 \Xi_2}{4} \log(u) X_2^4
    \nonumber\\
    &+
    \frac{15 \Xi_2 u^{2}}{448} X_2^4
    \left(-1415 - 280 X_2 + 196 X_2^2\right)
    \nonumber\\
    &+
    \mathcal{O}\left(c^{-3}\right) 
    \bigg\}  + 1 \leftrightarrow 2
    \,,
\nonumber \\
\label{eq:Psi-dissipation-only}
   &=
   \frac{3 u^3}{128 \eta}
   \bigg\{ 
    \frac{25}{32} \bar{\Xi} 
    -\frac{75 }{32} \bar{\Xi} \log(u)
    -
    \frac{11295  u^2}{1792}\delta \bar{\Xi}
    \nonumber\\
    &
    \hspace{2cm}
    +
    \mathcal{O}\left(c^{-3}\right)
    \bigg\}
\end{align}
where the $\bar{\Xi}$ and $\delta \bar{\Xi}$ are the binary tidal dissipative deformabilities defined in Eq.~\eqref{eq:xibar-delta-xibar}.
Observe that one of the leading order terms in the dissipative tidal phase of Eq.~\eqref{eq:Psi-dissipation-only} (the one that is independent of the $\log{(u)}$) contributes in the same way as the time of coalescence. Absorbing this term in a re-definition of $t_{c}$ via 
\begin{align}
    \bar{t}_c &\equiv t_c + \frac{75 M \bar{\Xi}}{8192 c^3 \eta}\,,
\end{align}
we then obtain the final result
\begin{align}\label{eq:GWphase-binary-final}
    \Psi\left(f\right)
    &=
    2\pi f \bar{t}_c
    -
    \varphi_c
    -
    \frac{\pi}{4}
    +
    \Psi_{\mathrm{pp}} +
    \Psi_{\mathrm{cons}}
    \nonumber\\
    &-
    \frac{3 u^3}{128 \eta}
    \bigg\{ 
    \frac{75 }{32} \bar{\Xi} \log(u)
    +
    \frac{11295  u^2}{1792}\delta \bar{\Xi}
    +
    \mathcal{O}\left(c^{-3}\right)
    \bigg\}\,.
\end{align}
We see then clearly that the $\log(u)$ term in the dissipative tidal correction to the Fourier phase ensures that the leading PN order correction is not degenerate with $\bar{t}_c$.
The leading PN order contribution to the phase due to dissipative tidal corrections was provided in Eq. (43) of~\cite{Ripley_2023}, and the above expression [Eq.~\eqref{eq:GWphase-binary-final}] provides the 1PN correction for the first time. 

\section{Analysis of GW170817}\label{sec:GW170817}
In this section, we analyze the gravitational wave event GW170817 by adding the next-to-leading PN order dissipative tidal correction to the gravitational wave phase.
The analysis with the leading order tidal correction was presented in~\cite{Ripley:2023lsq}, where we discussed the data analysis method in detail.
We review the main steps for completeness below and present our results using the 1PN corrected Fourier phase.

For the waveform model, we enhance the \texttt{IMRPhenomPv2\_NRTidal} model by adding the dissipative tidal correction [Eq.~\eqref{eq:Psi-dissipation-only}] to the gravitational wave phase.
In the frequency domain, we represent the model via the Fourier transform of the GW strain $\tilde{h}\left(f\right)= A\left(f;\theta\right) e^{i\Psi\left(f;\theta\right)}$, where $A(f;\theta)$ is the Fourier GW amplitude and $\Psi(f;\theta)$ is the Fourier GW phase. 
The parameters in our model are denoted by $\theta$ and the parameters in \texttt{IMRPhenomPv2\_NRTidal} are denoted by $\theta_a$. 
The enhanced phase $\Psi(f;\theta)$ is given by
\begin{align}
\label{eq:dissipative-tidal-phase-contribution}
    \Psi(f;\theta) = \Psi_{{\rm Pv2NRT}}(f;\theta_{a}) 
    +
    \Psi_{\mathrm{diss}}\left(f;\Bar{\Xi}, \delta \Bar{\Xi}\right)
    \,,
\end{align}
where the dissipative tidal contribution is provided in Eq.~\eqref{eq:Psi-dissipation-only}.
There are 19 parameters in our enhanced model;
the \texttt{IMRPhenomPv2\_NRTidal} GW model contains 17 parameters and we have included 2 dissipative tidal parameters. 
To sample the conservative tidal deformabilities we use the binary Love relations~\cite{Yagi:2015pkc}, and we marginalize over their uncertainty~\cite{Carson:2019rjx}.
We use these relations to rewrite $\lambda_a = (\Lambda_2 - \Lambda_1)/2$ as a function of $\lambda_s = (\Lambda_1 + \Lambda_2)/2$, thus reducing the total number of parameters to $17+2-1=18$. 

As usual in GW data analysis, we assume the noise is Gaussian and stationary, so that the log-likelihood of the strain data $\tilde{s}(f)$ given a GW template $\tilde{h}(f;\theta)$ with model parameters $\theta$ is given by~\cite{Maggiore-vol-1} 
\begin{align}
    \ln \mathcal{L}  (\tilde{s}| \theta)
    &= - \dfrac{1}{2}\left(
                \tilde{r}(\theta)|
                \tilde{r}(\theta)
            \right) 
    =
    -
    2 \int_0^{\infty} df 
        \frac{
        \left|\tilde{r}(f;\theta\right|^2          
        }{
            S_n\left(f\right)
        }
        ,
\end{align}
where $\tilde{r}(f;\theta) \equiv \tilde{h}\left(f;\theta\right) - \tilde{s}\left(f\right) $ is the residual and $S_n\left(f\right)$ is the noise power spectral density of the GW detector. We use 128s of the publicly available 4kHz GW170817 (glitch-cleaned) GW strain data \cite{LIGOScientific:2019lzm} for our data analysis. For sampling the likelihood, we use the \texttt{Bilby} \cite{Ashton:2018jfp} GW library with the nested sampling algorithm, as implemented in \texttt{DYNESTY} \cite{2020MNRAS.493.3132S}. 
Within the \texttt{Bilby} interface to that code, we set \texttt{nlive}$=1500$, \texttt{nact}$=10$, \texttt{dlogz}$=0.01$, \texttt{sample}$=$`\texttt{rwalk}', and \texttt{bound}$=$`\texttt{live}',
which has been verified to give convergent solutions~\cite{Ripley:2023lsq}. We sample the likelihood over all $18$ parameters of the model and we marginalize over the reference phase. 

We choose the following priors for our parameter estimation analysis. 
We derive the inferred prior on the chirp mass and mass ratio $q$ using uniform priors on the component masses.
We constrain the chirp mass $\mathcal{M}$ to lie within the range $\left[1.184\mathrm{M}_{\odot},1.25\mathrm{M}_{\odot}\right]$, and we constrain the mass ratio $q$ lies within the range $\left[0.5,1\right]$.
As the \texttt{IMRPhenomPv2\_NRTidal} model does not include spin corrections to the conservative tidal effects\footnote{We note that our dissipative tidal term contains no spin corrections either.}, we use the ``low-spin'' prior defined in \cite{LIGOScientific:2018hze}; that is, we use uniform priors for the neutron star spins $(a_{1},a_{2})$ in the range $[0,0.05]$. 
We use a triangular prior for $\lambda_s$ with mean $1500$ and range $\left[0,3000\right]$.
To sample the dissipative tidal deformabilities, we use uniform priors on the individual dissipative tidal deformabilities $(\Xi_{1},\Xi_{2})$ in the range $[0,8000]$.
The lower edge of this prior is set to zero because we exclude the possibility of anti-dissipative processes within each star ($\Xi_{1,2}<0$).
The upper edge of the prior is set by a heuristic constraint on the timescale for causal momentum transport across the star: dissipative/viscous effects should not transport momentum faster than the speed of light~\cite{Ripley_2023}. 
The rest of our waveform parameter priors follow the choices of \cite{LIGOScientific:2018hze,Ripley:2023lsq}.

The results of our Bayesian parameter estimation study on all the non-tidal parameters are consistent with~\cite{LIGOScientific:2018hze}.
We present the marginalized posteriors and priors on the dissipative tidal parameters $\Xi_{1,2}$ in Fig.~\ref{fig:posterior} (the full corner plots are included in Appendix~\ref{appendix:corner-plot}).
\begin{figure}[h!]
    \centering
    \includegraphics[width = 0.95\columnwidth]{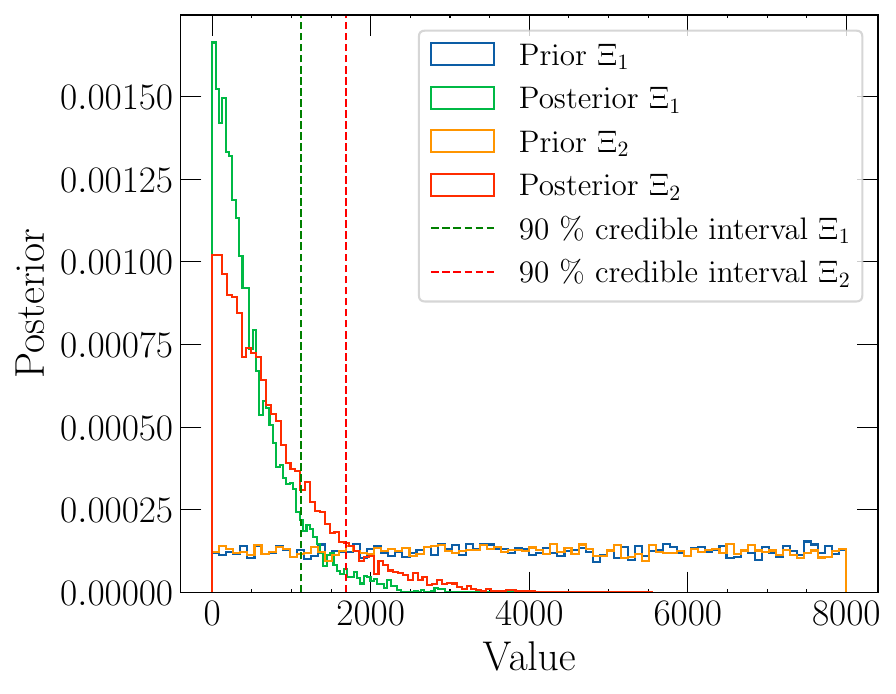}
    \caption{Marginalized posterior distribution for $\Xi_{1,2}$ for the GW170817 event and $90 \%$ credible intervals. 
    The priors for $\Xi_{1,2}$ are plotted in blue and orange respectively. 
    The marginalized posterior distribution of $\Xi_{1,2}$ are plotted as red and green histograms.
    We also show the $90\%$ credible interval for $\Xi_{1,2}$ as a green dotted line ($\Xi_1 \approx 1121$) and a red dotted line ($\Xi_2 \approx 1692$).
    Observe that the data is sufficiently informative to constrain the dissipative tidal parameters.}
    \label{fig:posterior}
\end{figure}
This figure shows that we can place the first-ever constraints on the individual tidal deformabilities of neutron stars using the GW170817 event: the data is sufficiently informative to constrain the dissipative tidal parameters to $\Xi_1 \leq 1121$ and $\Xi_2 \leq 1692$ at $90 \%$ credible levels. These constraints are consistent with the heuristic study we carried out previously in~\cite{Ripley:2023lsq}, where we used a uniform prior on $\Bar{\Xi}$ and assumed GW170817 was produced by an exactly equal-mass neutron star binary to obtain $\Xi_{1,2} \leq 1200 $ at $90 \%$ confidence. In this paper, we have improved on this heuristic estimate by using the next-to-leading order correction to the waveform and appropriately carrying out a Bayesian parameter estimation study.

Figure~\ref{fig:posterior} also shows that $\Xi_1$ is better constrained than $\Xi_2$ because of the dependence of $\bar{\Xi}$ and $\delta \bar{\Xi}$ on $\eta$.
Suppose that we have a system where $M_1 \gg M_2$.
Expanding $\bar{\Xi}$ and $\delta \bar{\Xi}$ in a small mass ratio expansion, we find
\begin{subequations}
\begin{align}
   \bar{\Xi} &= 
   \left(-8 \eta^4+16 \eta^2-32 \eta+8\right) \Xi_1
   +
   8 \eta^4 \Xi_2
   +
   \mathcal{O}\left( \eta^5\right)\,,\\
    \delta \bar{\Xi} &=
    \bigg(-\frac{5660 \eta^4}{753}+\frac{1568 \eta^3}{753}+\frac{3288 \eta^2}{251}-\frac{23536 \eta}{753}
    \nonumber\\
    &\hspace{1cm}
    +\frac{5996}{753}\bigg) \Xi_1
    +
    \frac{5660 \eta^4 \Xi_2}{753}
    +
    \mathcal{O}\left( \eta^5\right)
    \,.
\end{align}
\end{subequations}
We see that when $\eta \to 0$, both $\bar{\Xi}$ and $\delta \bar{\Xi}$ are both proportional to $\Xi_1$, and the dependence on $\Xi_2$ shows up at $\mathcal{O}(\eta^4)$.
Therefore, we see that $\Xi_1$ is the dominant contribution to both $\bar{\Xi}$ and $\delta \bar{\Xi}$.
Unless, there is prior information on $\Xi_{1,2}$, one will always constrain the dissipative tidal deformability of the larger mass better.
\section{Conclusions}\label{sec:conclusions}
In this paper, we have calculated the 1PN correction to the equations of motion and the  gravitational wave phase of a binary system in a quasi-circular inspiral undergoing dissipative tidal interactions. In our calculation, we truncated the tidal dynamics to electric quadrupolar contributions and derived the equations of motion using the Lagrangian method of~\cite{Vines_flanagan_2010}.
We then formulated a generalized energy-balance law for the dissipative tidal interaction.
Finally, we calculated the gravitational wave Fourier phase for a non-spinning binary system in a quasi-circular inspiral using the stationary phase approximation.
The final expression for the gravitational wave phase can be found in Eq.~\eqref{eq:GWphase-binary-final}.
We used this waveform to constrain the individual dissipative tidal deformabilities of each star using the GW170817 data through a Bayesian analysis.
The marginalized posteriors derived from our analysis are shown in Fig.~\ref{fig:posterior}, which present the first constraints on the individual tidal deformabilities of neutron stars. 
This analysis improves previous work that derived constraints on a certain combination of the individual tidal deformabilities by working with a waveform with leading PN order dissipative tidal effects. 

The work carried out here shows in detail that each objects dissipative tidal deformability, if present, imprints onto the gravitational waves emitted by inspiraling neutron stars in such a way that could potentially be observed in current and future gravitational wave events. Therefore, it is crucial that future work study the aspects of nuclear physics that one is constraining or measuring by inferring the value of the individual tidal deformabilities. In previous work, we have shown how to calculate these quantities, given an equation of state and the shear or bulk viscosity profile inside a neutron star. The latter, however, requires microphysical calculations that are only now becoming available. By combining these microphysical calculations with relativistic fluid and PN calculations, one should be able to determine precisely how the tidal deformabilities depend on the equation of state and out-of-equilibrium processes inside neutron stars, work that is currently ongoing. 

The conservative tidal deformabilities have been shown to satisfy approximately equation-of-state insensitive (binary Love) relations, which can be used to aid parameter estimation, so one may wonder if similar relations also exist for the dissipative tidal deformabilities. Future work could study the existence of such relations, although we suspect that perhaps the dissipative tidal deformabilities will not be as insensitive to the other important physics that control out-of-equilibrium processes in neutron stars, such as internal temperature.   

Another interesting line of future work is the study of the impact of dissipative processes in dynamical fluid excitations inside neutron stars. Previous work has shown that certain certain fluid modes may be excited into effectively simple harmonic motion, very close but before the merger of neutron stars, as the frequency of the fluid modes becomes near resonant with the orbital frequency~\cite{Hinderer_2016,Steinhoff_2016,Pitre:2023xsr,Pratten:2021pro}. One could imagine extended these calculations to include dissipative effects, which we expect will damp these dynamical excitations. Whether the effect of the dynamical excitations will remain or not will depend on whether the dissipative timescales in play are larger or smaller than the orbital time scale very close to merger. This, in turn, will depend on the microphysical processes that lead to dissipation in the first place, all of which is ripe for future study.  

Recently, Ref.~\cite{Chia:2024bwc} constrained 
the dissipative tidal effects of spinning black holes and exotic compact objects using data from the first observing run to third observing run of the Ligo-Virgo-Kagra collaboration~\cite{KAGRA:2021vkt}.
This reference only incorporated leading PN order, electric and magnetic quadrupolar spin-dependent effects using effective field theory techniques, in addition to the leading PN order, non-spinning contribution, when analyzing gravitational wave data.
It would interesting to see how the 1PN correction calculated in this paper improves their constraints on black hole tidal heating.
\begin{figure*}[thp!]
    \centering
    \includegraphics[width = 0.95\textwidth]{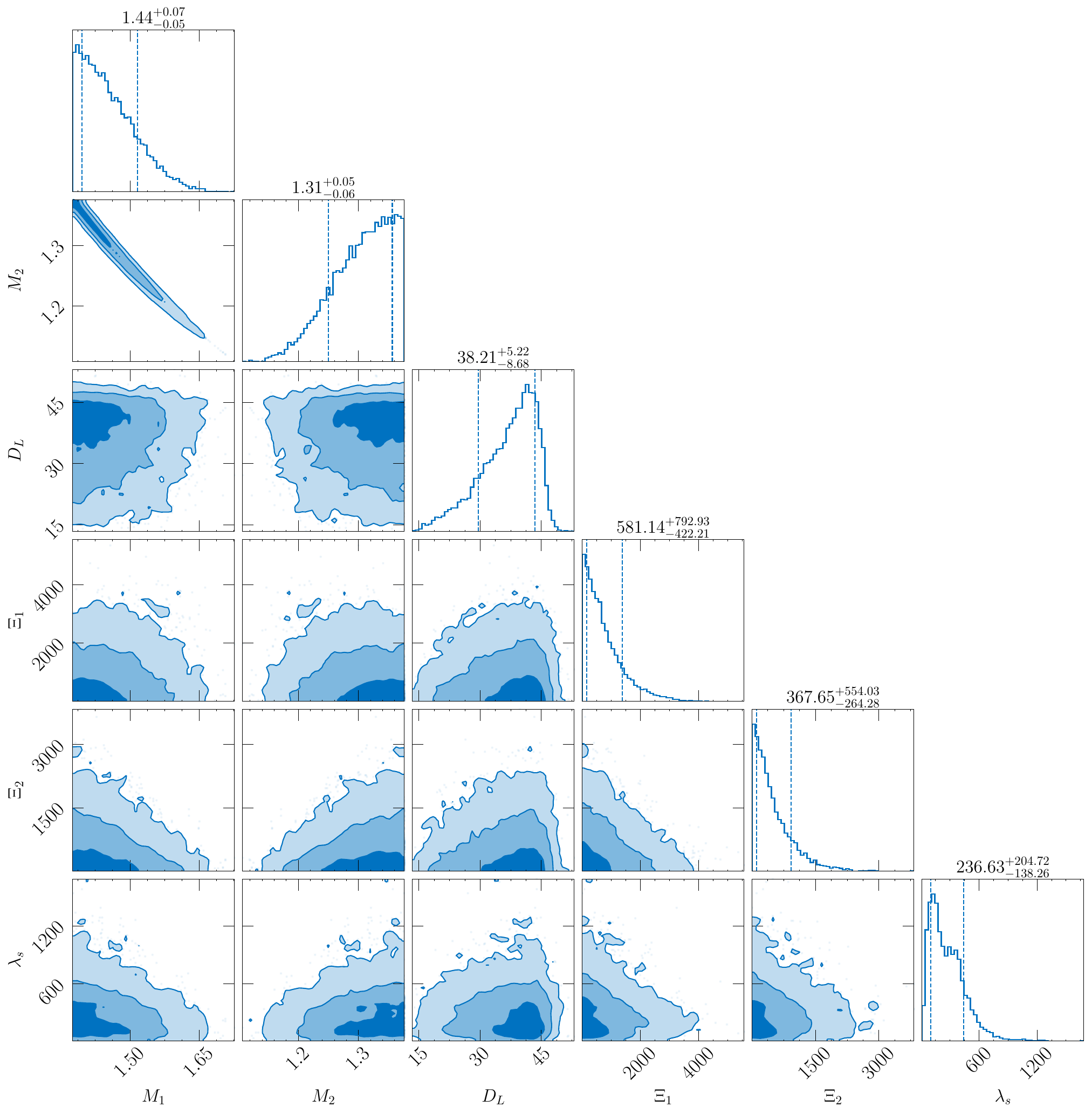}
    \caption{Corner plot from our analysis of GW170817 with next-to-leading order dissipative tidal corrections parameter space for a subset of parameters $(M_1, M_2, D_L, \Xi_1, \Xi_2, \lambda_s)$.}
    \label{fig:corner_plot}
\end{figure*}
\begin{acknowledgements}
We acknowledge support from the Simons Foundation through Award No. 896696, the National Science Foundation (NSF) Grant No. PHY-2207650, and the National Aeronatucis and Space Agency (NASA) Award No. 80NSSC22K0806.

\end{acknowledgements}
\appendix
\input{appendix}
\bibliography{ref}
\end{document}

%% file: appendix.tex
\appendix
\section{Values of \texorpdfstring{$B_i$}{}}\label{appendix:Bi-coeffs}
The values of the coefficients $B_i$ that appear in Eq.~\eqref{eq:aiQ-eqn} are given below:
\begin{align}
    &B_1=-\frac{15}{2X_2}(1+3\eta)\,,
    B_2=\frac{105X_1}{4} \,,
    \nonumber\\
    &
    B_3=- \frac{3 (-40 + 3 X_2 + 13 X_2^2)}{2 X_2}\,,
    \nonumber\\
    &B_4=\frac{3}{X_2}(2+2X_2-3X_2^2)\,,
    B_5=-\frac{15}{2X_2}(2-X_2-X_2^2)\,,
    \nonumber\\
    &B_6=-\frac{3}{X_2}(8-X_2-3X_2^2)\,,
    B_7=\frac{15}{X_2}(2-\eta)\,,
    \nonumber\\
    &B_8=-\frac{3}{2X_2}(7-2X_2+3X_2^2)\,,
    B_9=-\frac{15X_1}{2X_2}(1+X_2)\,,
    \nonumber\\
    &B_{10}=\frac{3X_1}{2X_2}\,,
    B_{11}=\frac{3}{2X_2}(5-4X_2-X_2^2)\,,
    \nonumber\\
    &
    B_{12}=-\frac{3}{2X_2}(4-X_2)\,,
    B_{13}=-\frac{15X_1}{2}\,,
    \\
    &
    B_{14}=\frac{6}{X_2} \,,
    B_{15}=-\frac{3X_1}{X_2} \,,
    B_{16}=\frac{3}{X_2}(1-2X_2-X_2^2) \,,
    \nonumber\\
    &
    B_{17}=\frac{3}{4} \,,
    B_{18}=\frac{3}{2} \,.
\end{align}
Note that all $B_{i}$, except $B_3$, are identical to those provided in Eq. (5.9 e) of VF~\cite{Vines_flanagan_2010}, although we do not include $B_{19}$, as this does not enter our final expressions. 
The difference arises because the tidal acceleration of VF $a^{i}_{Q,VF}$ is related to Eq.~\eqref{eq:aiQ-eqn} via
\begin{align}
    a^{i}_{Q,VF} = a^i_Q - \frac{3}{r^2 c^2} U_Q n^i -\frac{E_{2,\mathrm{int}}}{r^2 c^2}\,.
\end{align}
The reason for this difference is the splitting of the mass $M_2$ [Eq.~\eqref{eq:M2-split-VF}] used by VF.

\section{Values of \texorpdfstring{$A_i$}{}}\label{appendix:Ai-coeffs}
The values of the coefficients $B_i$ that appear in Eq.~\eqref{eq:LQ-eqn} are given below:
\begin{align}
    &A_1=\frac{3X_1 (3+\eta)}{4}\,,
    A_2=\frac{15 \eta X_1}{4} \,,
    \nonumber\\
    &
    A_3=- \frac{3X_1}{2}(1+3X_1)\,,
    A_4=\frac{3 X_1^2}{2}\,,
    \nonumber\\
    &
    A_5=-\frac{3 X_1^2}{2}(3+X_2)\,,
    A_6=-\frac{3 \eta}{2}\,,
    \nonumber\\
    &
    A_7=-\frac{3 \eta}{4}\,,
    A_8=\frac{X_1^2}{2}\,,
    A_9=X_1\,.
\end{align}
\section{Values of \texorpdfstring{$\alpha_i$}{}}\label{appendix:alphai-coeffs}
The values of the coefficients $\alpha_i$ that appear in Eq.~\eqref{eq:FdissQ-eqn} are given below:
\begin{subequations}
\begin{align}
\alpha_{1} &=
\tfrac{15}{4} X_1 (3 -  X_1 + 7 X_1^2)
\,, \\
\alpha_{2} &=
\tfrac{105}{4} (1 -  X_1) X_1^2
\,, \\
\alpha_{3} &=
- \tfrac{3}{2} X_1 (-9 - 17 X_1 + 2 X_1^2)
\,, \\
\alpha_{4} &=
- \tfrac{3}{2} (-7 + X_1) X_1^2
\,, \\
\alpha_{5} &=
- \tfrac{9}{2} X_1 (1 -  X_1 + 2 X_1^2)
\,, \\
\alpha_{6} &=
\tfrac{15}{2} X_1^2 (-5 + 2 X_1)
\,, \\
\alpha_{7} &=
\tfrac{3}{2} X_1 (-2 - 8 X_1 + X_1^2)
\,, \\
\alpha_{8} &=
- \tfrac{3}{2} X_1 (3 + X_1 + 2 X_1^2)
\,, \\
\alpha_{9} &=
\tfrac{15}{2} (-1 + X_1) X_1^2
\,, \\
\alpha_{10} &=
-3 X_1^2
\,, \\
\alpha_{11} &=
-3 (-4 + X_1) X_1^2
\,, \\
\alpha_{12} &=
- \tfrac{3}{4} (-1 + X_1) X_1
\,, \\
\alpha_{13} &=
- \tfrac{3}{2} (-1 + X_1) X_1
\,.
\end{align}
\end{subequations}
\section{Corner plot}\label{appendix:corner-plot}
We present the corner plot for subset $(M_1, M_2, D_L, \Xi_1, \Xi_2, \lambda_s)$ in Fig.~\ref{fig:corner_plot}.
We have verified that our analysis is consistent with that of~\cite{LIGOScientific:2018hze} for the non-dissipative tidal parameters and the non-tidal parameters.